\documentclass[sigconf]{acmart}
\usepackage{float}
\setcopyright{none}
\renewcommand\footnotetextcopyrightpermission[1]{}

\begin{document}

\title[The Path to Conversational AI Tutors]{The Path to Conversational AI Tutors: Integrating Tutoring Best Practices and Targeted Technologies to Produce Scalable AI Agents}
% For every author affiliation, add \country{...}

\author{Kirk Vanacore}
\email{kpv27@cornell.edu}
\affiliation{%
  \institution{Cornell University}
  \country{USA}
}

\author{Ryan S. Baker}
\email{ryanshaunbaker@gmail.com}
\affiliation{%
  \institution{University of Adelaide}
  \country{Australia}
}

\author{Avery H. Closser}
\email{avery.closser@ufl.edu}
\affiliation{%
  \institution{University of Florida}
  \country{USA}
}

\author{Jeremy Roschelle}
\email{jroschelle@digitalpromise.org}
\affiliation{%
  \institution{Digital Promise}
  \country{USA}
}

\begin{abstract}
The emergence of generative AI has accelerated the development of conversational tutoring systems that interact with students through natural language dialogue. Unlike prior intelligent tutoring systems (ITS), which largely function as adaptive and interactive problem sets with feedback and hints, conversational tutors hold the potential to simulate high-quality human tutoring by engaging with students’ thoughts, questions, and misconceptions in real time. While some previous ITS, such as AutoTutor, could respond conversationally, they were expensive to author and lacked a full range of conversational ability. Generative AI has changed the capacity of ITS to engage conversationally. However, realizing the full potential of conversational tutors requires careful consideration of what research on human tutoring and ITS has already established, while also unpacking what new research will be needed. This paper synthesizes tenets of successful human tutoring, lessons learned from legacy ITS, and emerging work on conversational AI tutors. We use a keep, change, center, study framework for guiding the design of conversational tutoring. We argue that systems should keep proven methods from prior ITS, such as knowledge tracing and affect detection; change how tutoring is delivered by leveraging generative AI for dynamic content generation and dialogic scaffolding; and center opportunities for meaning-making, student agency, and granular diagnosis of reasoning. Finally, we identify areas requiring further study, including efficacy testing, student experience, and integration with human instruction. By synthesizing insights from human tutoring, legacy ITS, and emerging generative AI technologies, this paper outlines a research agenda for developing conversational tutors that are scalable, pedagogically effective, and responsive to the social and motivational dimensions of learning.
\end{abstract}




\maketitle

\section{Introduction}
Generative Artificial Intelligence’s (AI) capacity to engage students through dynamic conversations is transforming intelligent tutoring systems (ITS). Though a few prior systems successfully deployed dialogue interaction systems (e.g., AutoTutor, Watson Tutor), many more digital learning platforms now incorporate Gen AI features that allow students to converse with AI tutors. Whereas the majority of current ITS systems function, in essence, as interactive and adaptive problem sets, conversational tutors may be able to simulate high-quality human tutors by responding to students’ thoughts, queries, and answers in real time. This emerging class of ITS can be referred to as \textit{conversational AI tutoring—systems} that engage learners in natural language dialogue, rather than digital worksheets or scripted problem sets.

Conversational AI tutors have the potential to overcome many limitations of earlier systems by enabling deep engagement with concepts while addressing students’ social-emotional and motivational needs. However, to be maximally effective, their interactions must be guided by evidence from human tutoring research and from the most successful technological innovations found in previous generations of ITS. Today, conversational tutors can be easily and quickly built, and many current examples show considerable promise for benefiting learning. Yet they also tend to rely on the default behaviors of large language models (and their system prompts), often only augmented by well-crafted prompts (e.g., \cite{Kestin2025Outperforms}) and sometimes fine-tuning \cite{team2025aiEedi}. Much of this effort is aimed at getting the pedagogy right; to do so requires understanding what good tutor pedagogy is and how it can be instantiated within systems that are flexible enough to respond to students’ needs at any moment and capable of supporting each student's sustained progress on challenging educational goals. 

However, this effort has largely not paid attention to what was learned and developed in the previous generations of intelligent tutors. Done well, conversational tutors could offer the ''best of both worlds:'' the subtle responsiveness of expert human tutors alongside the systematic support and precision of established ITS technologies (e.g., knowledge tracing, automated feedback, behavior detection). In order to support the field in collectively achieving this goal, we present a review of relevant research on expert human tutoring and previous-generation ITS. Then we propose a \textit{keep}, \textit{change}, \textit{center}, \textit{study} framework: identifying what to \textit{keep} from prior ITS, what to \textit{change} in light of new capabilities, what new opportunities to \textit{center} as part of the student’s AI tutor experience, and what must be \textit{studied} further to understand how students respond to conversational tutoring. We believe this approach will bring us closer to the goal of scalable AI tutoring systems that model high-impact human tutors.

\section{What Makes Conversational Human Tutoring Effective?}
Human tutoring is widely recognized as one of the most effective educational interventions, providing a model that many have strived to emulate in the design of ITS \cite{VanLehn2011, Kraft2020}. Seminal studies demonstrate that tutored students consistently outperform their peers and report more positive attitudes toward the subject \cite{Cohen1982, Slavin1987, Kraft2020}. In a comprehensive review of educational interventions, Kraft \cite{Kraft2020} demonstrated that tutoring programs consistently outperform common alternatives such as class-size reduction and summer school, yielding effect sizes that significantly exceed the average impact of typical school-based programs.

However, despite its success, scaling quality one-on-one tutoring remains a persistent challenge due to costs, implementation consistency, and logistical barriers \cite{KraftFalken2021}. Furthermore, many tutoring programs experience a decrease in impact as they scale, ostensibly due to the lack of quality tutoring that comes with large-scale programs \cite{Kraft2024}. Many tutors also receive little training in best practices, making it difficult to ensure students consistently experience high-quality instruction \cite{Cohen1982, Chi2001}. Efforts to implement tutoring programs with fidelity and integrate them into school schedules require significant resources, and out-of-school programs often face additional challenges of attendance and quality \cite{KraftFalken2021}.

These constraints make it hard to scale human tutoring, but they do not diminish the importance of understanding \textit{why it works}. The effectiveness of tutoring lies not only in individualized attention but also in the conversations themselves: moments where tutors guide thinking, clarify meaning, and build relationships that sustain motivation. Decades of research converge on three dimensions that explain why tutoring conversations are effective: pedagogical techniques that guide learning, methods that sustain motivation and engagement, and interpersonal relationships that build trust.

\subsection{Pedagogical Techniques}
One of the clearest lessons from research on human tutoring is that conversations do more than deliver explanations and answers: they scaffold students’ reasoning. Most tutoring conversations involve students working through content while tutors provide feedback, scaffold problem-solving, and model strategies. One of the key benefits of tutoring over independent work is the immediacy and specificity of feedback \cite{Shute_2008}. High-quality tutors go beyond feedback on accuracy, often guiding students as they think through errors. Graesser et al. \cite{Graesser1995} refer to this guidance as a “scaffolding episode,” which involves iterative steps guiding students toward correct answers or problem-solving processes.

A central dimension of effective scaffolding is subtlety. Expert tutors often provide indirect cues that prompt students to self-correct their mistakes \cite{Merrill1992, Fox1991}. These methods of ``cueing'' tend to have higher impacts on learning outcomes than other forms of feedback \cite{Hattie2007}. To do this, tutors often present leading questions or more general prompts without directly referring to the content (e.g., “what else?”, “interesting choice”) \cite{Chi2001, McArthur1990}. These conversational moves help ensure that students, rather than tutors, do the majority of the cognitive work.

At the same time, subtle cues are not always enough. When students reach an impasse, expert tutors provide more direct scaffolding by controlling the elements of a problem that are currently beyond the student's ability, allowing them to focus on subcomponents that are within reach \cite{Wood1976}. Such scaffolding often includes demonstrations, such as worked examples or explanations \cite{Juel1996}. Research shows that explanations are more effective when integrated into scaffolded processes rather than presented as “prefabricated information” \cite{Wittwer2008}. Moreover, explanations are most effective when students self-explain and rephrase them \cite{Renkl1997, McKendree1990, Bielaczyc1995}, highlighting the importance of interactive dialogue over passive reception.

Effective tutoring conversations also foster metacognition by prompting students to reflect on their own reasoning. High-quality tutors guide students toward self-evaluation, a key component of self-regulated learning \cite{zimmerman2009self}. Chi et al. \cite{Chi2001} found that expert tutors elicited more robust responses to self-evaluation and self-explanation queries compared with lower-quality tutors, and these queries were associated with higher learning gains. Similarly, Graesser \& Person \cite{Graesser1994} found that students who produced higher-quality questions during tutoring sessions achieved greater learning gains than those who produced lower-quality questions. These findings suggest that effective tutors aren’t just providing students with correct information; they are cultivating habits of self-questioning and self-reflection through conversation.

\subsection{Engagement and Motivation}
Tutoring conversations don’t just deliver instruction; they also shape how students feel about learning, using dialogue to support motivation and engagement. Expert tutors can foster positive attitudes toward subject matter, stronger academic self-concept, social skills, and motivation \cite{Cohen1982,Merrill1992,BowmanPerrott2014}. Self-determination theory provides a useful lens here: students remain engaged when conversations support their core psychological needs for competence, autonomy, and relatedness \cite{RyanDeci2012}.

One key need is competence. Students are more motivated when they feel capable of mastering material. Effective feedback is strongly linked to greater motivation and engagement \cite{Hattie2007}. Lepper and Woolverton \cite{Lepper2002} found that tutors differentiated scaffolding to accommodate each learner while fostering motivation by highlighting small successes. Personalized feedback that helps students overcome misconceptions, combined with affirmation when they succeed, provides a motivational boost by reinforcing competence \cite{RyanDeci2012}.

A second mechanism is autonomy and agency. Researchers have shown that the way feedback is provided can influence students’ sense of control over learning. Subtle interventions can enhance students’ sense of autonomy, improving motivation \cite{Merrill1992, Fox1991, RyanDeci2012}. By giving students space to make decisions as they work through material, while offering encouragement and support, expert tutors can increase intrinsic motivation \cite{Lepper2002, Chi2014}. The ability to dynamically respond to a student’s unique problem-solving path allows the learner to take ownership of their progress in ways that are rare in other contexts.

Finally, motivation is supported through appropriate challenge, where learning activities fall within a student’s zone of proximal development, which captures activities that students can complete with support \cite{Vygotsky_Cole_1978}. Tutors help maintain students in this zone by adjusting the level of support so that tasks are neither too easy (which can reduce engagement) nor too difficult (which can lead to frustration). This dynamic calibration echoes Csikszentmihalyi’s theory of flow \cite{Csikszentmihalyi1990}, in which engagement is maximized when challenge and skill are in balance. When tutors provide problems within students’ zone of proximal development, learners are more likely to sustain interest and engagement \cite{Lepper2002}.

\subsection{Interpersonal Relationships}
Tutoring conversations are also relational: the quality of the tutor–student relationship directly shapes learning and motivation. When students experience a caring, warm, and supportive relationship with their tutor, they show higher motivation and achievement \cite{Wentzel1997}. Multiple studies highlight the importance of supportive tutoring relationships \cite{Juel1996,KraftFalken2021}, which can evolve into mentorship roles beyond specific academic interventions. Kraft and Falken \cite{KraftFalken2021} argue that these interpersonal bonds are so central to tutoring success that programs should intentionally integrate best practices from both mentoring and tutoring. Such relationships may underlie many of the broader social and motivational benefits of tutoring, as a meta-analysis found that tutoring positively affected social skills, behavior, and academic engagement — with the strongest effects emerging when rewards were not used \cite{BowmanPerrott2014}.

Rapport also enhances the pedagogical and motivational dimensions of tutoring. One study found that higher rapport between peer tutor dyads was associated with students completing more problems \cite{Sinha2016}. Another showed that tutors who established stronger rapport were more effective at eliciting metacognitive engagement, particularly by encouraging students to explain their reasoning \cite{Madaio2018}. These findings underscore that tutoring conversations are not purely instructional exchanges: they are social interactions where trust, rapport, and care create the conditions for deeper learning.

\subsection{What AI Tutors Need to Learn from Human Tutors}

Taken together, research on human tutoring indicates that its effectiveness depends on the quality of conversations tutors have with students. High-quality conversations do more than deliver instruction: they scaffold reasoning through well-timed pedagogical techniques, sustain motivation by meeting students’ needs for competence, autonomy, and appropriate challenge, and build relationships of trust and care that create the conditions for learning. In each case, dialogue is the medium through which these effects emerge — whether it is prompting a student to self-explain, affirming progress to boost competence, or cultivating rapport that deepens engagement.

For the design of conversational tutoring systems, these findings highlight that the goal is not simply to automate feedback or explanations. To be maximally effective, AI tutors will need to reproduce the qualities of human tutoring conversations that guide, motivate, and connect with students. Furthermore, they will need to scaffold students’ reasoning by providing support and allowing students the space to work challenges out for themselves. The challenge, then, is to design systems that can sustain these kinds of interactions at scale, preserving the conversational richness that makes human tutoring so powerful. At the same time, some students may respond differently to the exact same text if it comes from an AI rather than a person, so different strategies may sometimes be needed to achieve these goals.

\section{Emerging Generation of Conversational Tutors}
The recent increase in available foundational generative AI models has inspired numerous new educational technology features and platforms. Although a comprehensive review of all current generative AI applications in ITS is beyond the scope of this paper, an overview of major developments and research areas provides helpful context for future directions. Current research and development primarily fall into two categories. First, many non-conversational generative AI features have been integrated into the current generation of ITS. generative AI is being increasingly used for evaluating student work, generating instant feedback, and increasing personalization within existing digital learning platforms  \cite{usher2025generative,nzenwata2024systematic}. Insights from research in these areas can inform the design of dialogue-based AI tutoring systems. Second, conversational tutoring systems are emerging in which students directly engage with generative AI outputs in real time during learning activities. Studies in both areas highlight current limitations and potential directions for future ITS advancements.

\subsection{Current Conversational AI Tutors}

Many researchers and educational technology companies are building, studying, and deploying generative AI tutors in a variety of applications with varying results. Perhaps the most prominent use of generative AI to tutor students is Khanmigo, produced by Khan Academy \cite{KhanAcademy2023}. This system offers instruction and support in math, science, and humanities. It has features like \textit{Tutor Me} that can generate problems based on what a student wants to learn and provide scaffolding, \textit{Refresh Me} that quizzes based on what topic a student inputs, and \textit{Write} that provides feedback on essays. Other conversational AI tutors provide a more limited focus. For example, Rori is a large language model-based math tutor run through WhatsApp, which allows students to work through predefined lessons on specific math topics through AI-generated chats \cite{Henkel2024}. Another system, \textit{OpenTutor}, provides lessons on very specific topics (e.g., Suicide Risk and Diode Bias) guided by an AI chat agent \cite{Nye2021}, and uses an LLM  for multi-concept short answer grading in order to guide students to correct solutions \cite{nye2023generative}. Other studies found that a conversational AI tutor could increase inclusion in higher education, especially in classes with very high student-to-teacher ratios \cite{gupta2022supporting}. 

Thus far, very few extensive evaluations or efficacy studies have been run on these systems to understand how they interact with learning processes and affect outcomes. In one randomized control trial, researchers found that a ``content-rich prompt engineered'' generative AI chat tutor improved performance and engagement in a Harvard Physics course \cite{Kestin2025Outperforms}. One qualitative evaluation of Khanmigo used for language learning found that although the system supported outcome focus and authentic learning experiences, it did not always provide the learner with tasks tailored to the learner's needs and ability, and it did not support students' metacognitive learning skills \cite{Shetye_2024}.  In a preliminary efficacy study, Henkel et al. \cite{Henkel2024} found that Rori had a moderate positive effect on student math skills. 

There is also relevant evidence that generative AI can be leveraged to enhance human tutoring interactions. This research stands as a kind of human-in-the-loop AI tutoring, in which humans review and edit AI-generated messages. In a large-scale randomized controlled trial, Wang et al. \cite{wang2024tutor} found that providing human tutors with real-time, LLM-generated pedagogical suggestions increased the probability of students mastering knowledge components by 4 percentage points on average. This effect was particularly pronounced for lower-performing or less-experienced tutors.  More recently, Google DeepMind partnered with Eedi to have LearnLM generate messages for chat-based tutoring sessions embedded into Eedi's adaptive learning platform \cite{team2025aiEedi}. These messages were reviewed by human tutors before being sent to the student. Over three-quarters of the messages generated by LearnLM did not need major revision by the human tutors. This suggests that while generative AI holds promise for independently interacting with students, even fine-tuned models may be insufficient to consistently determine what instruction a student needs without human oversight. Another study found that fine-tuning models with educational data enabled generative AI tutors to provide personalized instruction; however, this was based on their interactions with synthetic students \cite{Perczel_Chow_Demszky_2025}.

There is also a growing number of benchmarking systems meant to evaluate LLMs instructional ability. For example, \textit{AI-for-Education.org} evaluates AI models based on their performance on pedagogy exams \cite{Leli_2025}. However, recent research suggests that even top-performing models on this benchmark have difficulty interpreting \textit{en vivo} educational dialogue \cite{vanacore2025well}, suggesting that performance on pedagogy exams may not translate to real-world educational scenarios. Google Research also released an evaluation of five generative AI systems for learning; however, this evaluation only compared expert tutors' preferences for different models, not the models’ impact \cite{jurenka2024towards}.

\subsection{Non-Conversational Applications}
Generative AI is also being used to create non-conversational educational tools, the capabilities of which may be leveraged towards conversational AI tutors in the future. New tools utilizing generative AI are being developed for grading open-response questions across various subjects (e.g., generative AI Smart Grading, \cite{tobler2024smart}; Gradescope, \cite{hansel2024gradescope}). Generative AI models produce promising results when grading mathematical open-response questions  \cite{Henkel2025}, short-answer questions \cite{floden2024grading,liu2024ai_assisted}, reflective essays \cite{Awidi2024}, academic writing tasks \cite{pack2024large}, and comprehension assessments \cite{henkel2023comprehension}. To achieve accuracy and consistency comparable to human graders, researchers have employed multi-shot prompting, chain-of-thought prompting, and detailed rubrics. Iterative design and tuning of these generative AI grading systems are required. However, even after using these methods to cull hallucinations, generative AI does not achieve perfect accuracy in mathematics problems \cite{Henkel2025}, nor is it always consistent with human grading \cite{floden2024grading}.

Although generative AI occasionally produces errors, multiple studies demonstrate that it can effectively create educational content, hints, and feedback when used in controlled environments. For instance, generative AI-generated hints are as effective as human-generated hints at improving student performance in mathematics \cite{Pardos_Bhandari_2024,Worden_Vanacore_Haim_Heffernan_2025}.  In computer science education, generative AI assists students in debugging code, provides explanations, generates tasks, and even grades student submissions \cite{pankiewicz2023large,lagakis2023automated}. These generative AI applications typically perform comparably to human equivalents \cite{lagakis2023automated,Worden_Vanacore_Haim_Heffernan_2025}. However, there is evidence that over-reliance on generative AI might negatively affect students' self-regulated learning behaviors, potentially impacting their ability to retain and transfer knowledge to novel contexts \cite{viberg2025chatting,vanacore2025unpacking}.

Additionally, the effectiveness of many generative AI-generated materials depends on rigorous quality review processes. Pardos and Bhandari \cite{Pardos_Bhandari_2024} required human experts to review GPT-3.5-generated hints before implementation, ultimately discarding between 25\% and 47\% of generated hints, depending on the topic. Conversely, Worden et al. \cite{Worden_Vanacore_Haim_Heffernan_2025} employed a self-verification procedure where GPT-4 both created and reviewed hints prior to use, identifying that 14\% of hints were inaccurate or inappropriate for students' grade level. Notably, human review did not find any additional errors. These findings suggest that while generative AI can produce effective educational content, ensuring content accuracy requires careful validation procedures. Beyond this, thoughtful implementation of generative AI systems is essential to foster productive learner behaviors rather than inadvertently encouraging superficial engagement.

\section{Framework for Conversational AI Tutors}
\begin{table*}[t]
  \caption{Summary of Keep, Change, Center, Study Framework for Conversational AI Tutors}
  \label{tab:framework}
  \begin{tabular}{p{1.5cm}p{4cm}p{4cm}p{5cm}}
    \toprule
    Category & Summary & Example(s) & Rationale \\
    \midrule
    \textbf{Keep} & Evidence-based ITS features that can be applied in new conversational AI tutors & Knowledge tracing, Behavior and affect detection, Knowledge Graphs  & These models curate content to keep students engaged and appropriately challenged, aligned with ZPD and flow. \\[5pt]

    \textbf{Change} & Potential of Generative AI in Tutoring & Automatic content generation, Dialogic feedback & Moves beyond fixed problem libraries; adapts instruction to the learner's background instantaneously. \\[5pt]

    \textbf{Center} & Opportunities Unique to Conversational AI Tutoring & Meaning-making, Student agency, Diagnosis of reasoning & Natural language enables the co-construction of meaning and granular assessment of reasoning. \\[5pt]

    \textbf{Study} & Research Directions for Conversational AI Tutoring & Efficacy testing, Student experience, Human-AI collaboration & Addresses misalignment with evidence-based practices; ensures systems remain trustworthy and scalable. \\
    \bottomrule
  \end{tabular}
\end{table*}

Taken together, these developments point to both the opportunities and limitations of current generative AI applications in tutoring. While current systems demonstrate that AI can generate content, provide feedback, and even sustain dialogue, they also underscore the importance of deliberate design choices to ensure accuracy, foster productive learning behaviors, and take advantage of evidence on what works coming from decades of tutoring research. To chart a path forward, we propose a framework for conversational AI tutors that draws on three complementary sources of insight: what to keep from legacy ITS, what to change in light of generative AI’s new affordances, and what to center as entirely new possibilities enabled by conversational interaction. Finally, we outline key areas the field must study to ensure these systems are effective, scalable, and aligned with human learning needs.

\subsection{Keep: Integrating Legacy ITS and Modern Conversational Design}
The move toward conversational AI tutors is not a departure from the past, but rather an evolution that must be grounded in decades of research and development. While legacy ITS were often domain-specific and highly structured, their development over the past several decades has produced a set of well-validated technologies that remain highly relevant for the next generation of conversational tutors. They laid the critical groundwork for delivering adaptive practice, precise feedback, and individualized pacing. Learning from this earlier generation provides both essential design patterns and cautionary notes for the current era of generative AI.

\subsubsection{The Inner–Outer Loop Architecture}

A defining conceptual advance of the legacy era was the inner–outer loop architecture \cite{VanLehn2006}. This dual structure enabled systems like Cognitive Tutor, ANDES, and ALEKS to provide finely tuned adaptivity both within and across problems. The inner loop managed step-level interactions within a single problem—detecting errors, providing correctness feedback, and offering graduated hints when students reached impasses. Conversely, the outer loop selected which problems to give students next, guided by knowledge tracing models and curricular sequencing rules \cite{Corbett1994}. The separation of these loops remains one of the most influential design patterns in the ITS tradition because it allows the system to guide students through appropriate questions. In generative AI conversational tutors, this architecture could be used to constrain the tutoring session to appropriate content within and between problems.

\subsubsection{Mapping the Domain: Knowledge Spaces and Graphs}

Beyond the architectural loops, legacy systems provided a robust methodology for representing the structure of knowledge through Knowledge Space Theory (KST) and Knowledge Graphs \cite{Peng_Xia_Naseriparsa_Osborne_2023}. Systems such as ALEKS demonstrated the power of using a knowledge space—a mathematical structure that maps all possible states of a student's knowledge based on the prerequisite relationships between concepts \cite{essa2016possible}. Knowledge space maps can help contribute to mapping students' misconceptions and lack of requisite knowledge \cite{pavlik2013review}. Modeling of misconceptions based on prior students' performance improves estimates of student knowledge that correlate to generalizable measures of learning \cite{Liu_Patel_Koedinger_2016}. Thus, connecting students' ``in-the-moment'' performance to specific misconceptions and lacking prerequisite knowledge may serve as a powerful tool assisting LLMs in generating targeted remedial content.

While AI can generate vast amounts of content, it often lacks a coherent "mental map" of how one concept logically leads to another. Integrating legacy knowledge graphs into the "outer loop" of conversational tutors allows for a more structured navigation of a domain. By grounding an LLM in a formal knowledge graph, the system can ensure that a conversational session does not leap to advanced concepts before the student has demonstrated mastery of foundational ones (e.g., multiplication before addition).  The recent Knowledge Graph release by Learning Commons provides an opportunity for integrating this important legacy feature directly into generative AI \cite{LearningCommons2024}. 

\subsubsection{Precision in Knowledge Modeling}

At the core of legacy ITS was the effort to model student cognition to approximate how learners acquire and apply knowledge. Early systems embedded production-rule cognitive models that mirrored hypothesized mental steps \cite{Anderson1995}, leading to the development of student knowledge modeling (“knowledge tracing”) algorithms. Bayesian Knowledge Tracing, for example, modeled mastery as a probabilistic state that evolved with practice \cite{Corbett1994}. This allowed systems to maintain specific "memories" of a student’s mastery—identifying exactly what they know and where they struggle. Later innovations, such as Performance Factor Analysis \cite{Pavlik2009} and Deep Knowledge Tracing \cite{Piech2015}, applied statistical and deep learning techniques to capture even more complex learning trajectories. More recent adaptations have allowed for advances such as how knowledge is tracked across systems with complex sequences of questions and optimizations attend to the relative importance of problems in estimating student knowledge \cite{Abdelrahman2023}.

The precision of these probabilistic models represents a central strength that conversational tutors must retain. While generative AI may have advantages in navigating the nuances of the ``inner loop'' dialogue, emerging evidence indicates that generative AI tutors currently struggle to infer student knowledge accurately, either at the outset of an interaction or as the activity progresses \cite{Shetye_2024,Vajjala_2026}. Integrating established knowledge tracing into conversational systems is therefore critical for ensuring that learners receive content aligned with their current level of understanding.

\subsubsection{Behavioral and Affective Modeling}
Previous-generation ITS provide advances in behavior and affect detection, developing models to identify when students were gaming the system \cite{Baker2008}, showing unproductive persistence ("wheel-spinning") \cite{Beck2013}, or experiencing affective states such as boredom, confusion, and frustration \cite{DMello2012Dynamics}. Although generative AI may have some native understanding of student behavior and affect, they were not trained for these specific purposes. Alternatively, the ITS behavior and affect detectors were developed using intensive in-person observational protocols \cite{ocumpaugh2015baker}, which notably requires data that was inaccessible for training the large language models that power generative AI (i.e., \textit{en vivo} observations). Thus, it is advantageous for newer tutors to lean on the data and methods used for these detectors to better understand and respond to students.

Although these ITS detectors achieved respectable accuracy, they were historically underutilized in real-time pedagogical responses—often logged for research rather than used to shape instruction. In the next generation of conversational tutors, these legacy detectors provide an empirical basis for improving responsiveness. Outputs from affect and behavior models can now be integrated into the conversational flow through prompt conditioning or retrieval-augmented generation (RAG), enabling AI tutors to respond with empathy and strategy when a student becomes frustrated or disengaged.

\subsection{Change: Potential of Generative AI in Tutoring}
While the integration of key technologies from previous-generation ITS technologies remains critical, the affordances of generative AI create opportunities to extend tutoring systems in ways that were previously infeasible. In particular, advances in content generation, conversational feedback, and motivational responsiveness illustrate how conversational tutors may move beyond the constraints of earlier systems (e.g., \cite{tobler2024smart,Pardos_Bhandari_2024,pankiewicz2023large}). At the same time, these opportunities must be balanced against the limitations of large language models, which pose challenges for educational deployment.

One promising area of change is bespoke content generation that aligns with students' needs and interests. Generative AI models are able to produce instructional material that is adapted to the learner’s background, preferences, or immediate needs almost instantaneously. This capacity extends far beyond the problem libraries of traditional ITS, allowing tutors to create novel practice opportunities, explanations, or examples in real time. Early implementations (e.g., Khanmigo) have already demonstrated the feasibility of tailoring content to student requests or to targeted areas of difficulty \cite{KhanAcademy2023}. If carefully designed, this capacity could make possible an effectively unlimited supply of individualized instructional resources, including for lower-demand topics that previously would have been economically infeasible to create content for.

A second area concerns feedback and scaffolding by probing student thinking through dialogue. Prior research has highlighted that much of human tutoring’s effectiveness derives from the ability to provide interactive feedback that encourages students to engage in constructive behaviors, such as self-repair and knowledge construction \cite{VanLehn2011, Chi2001}. While earlier ITS incorporated personalized feedback systems, including in natural language (e.g., \cite{graesser1999autotutor}), their responses were typically rule-based and limited in scope. Conversational tutors based on Generative AI, by contrast, have the potential to engage in open-ended dialogue that flexibly adapts to a student’s reasoning path. This may enable new forms of interactive scaffolding that approach, and in some cases exceed, the responsiveness of human tutors.

However, these possibilities are tempered by the current limitations of Gen AI. Large language models frequently produce inaccurate or fabricated content (“hallucinations”) \cite{Ji2023Hallucination}, exhibit tendencies toward over-helpfulness that may reduce students’ metacognitive effort \cite{Fan2025}, and often struggle to manage multi-turn dialogue that requires sustained pedagogical goals \cite{li2025beyond,Nye2021}. Moreover, as discussed above, they remain limited in their ability to accurately diagnose students’ knowledge states \cite{Shetye_2024}. For generative AI conversational tutors to realize their full potential, these limitations must be addressed through careful design and integration with complementary methods.

\begin{figure*}[!htbp]
    \centering
    \includegraphics[width=0.95\linewidth]{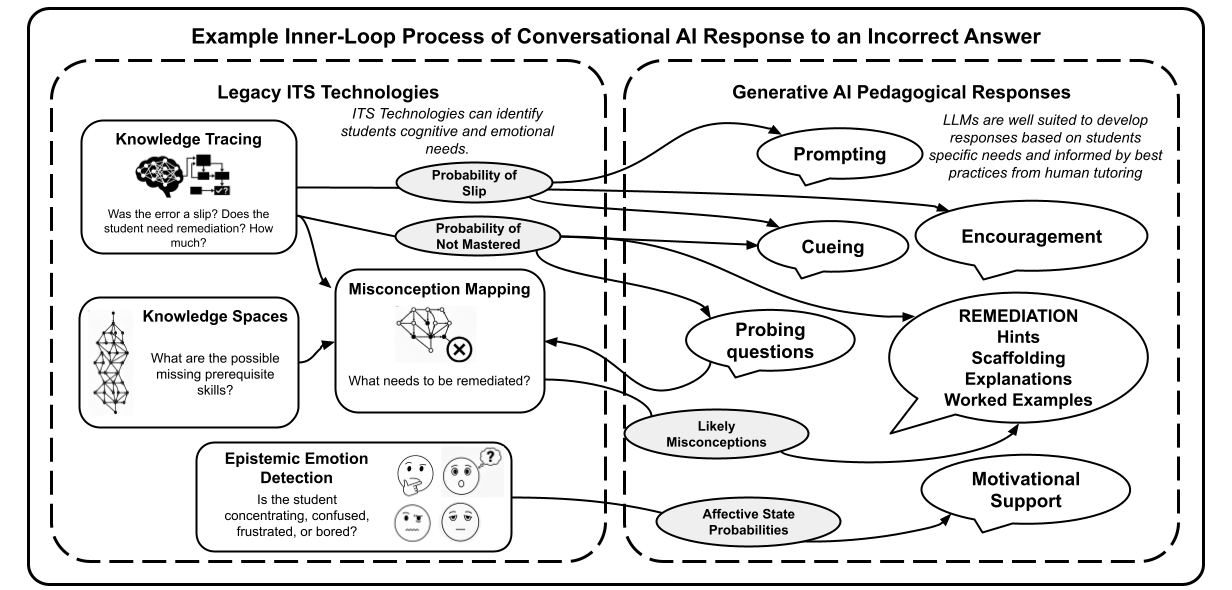}
    \caption{This diagram illustrates an Example Inner-Loop Process where legacy Intelligent Tutoring System (ITS) technologies are integrated with Generative AI to manage student errors. On the left, Legacy ITS Technologies—such as Knowledge Tracing, Knowledge Spaces, and Epistemic Emotion Detection—diagnose a student's cognitive and emotional state, identifying probabilities of "slips," lack of mastery, or specific misconceptions. These data-driven insights then inform the Generative AI Pedagogical Responses on the right, which leverage Large Language Models (LLMs) to deliver tailored interventions that are informed by established best practices in human tutoring.}
    \label{fig:diag}
\end{figure*}

\subsection{Center: Opportunities Unique to Conversational AI Tutoring}
Beyond retaining proven methods and extending existing capacities, conversational tutoring allows us to center new dimensions of learning that were difficult or impossible to realize in prior ITS. These include deeper engagement with meaning through dialogue, greater student agency in shaping learning interactions, and more nuanced diagnosis of reasoning at the conversational level.

One opportunity is to center meaning-making and conceptual dialogue. While many previous ITS focused primarily on procedural practice and correctness, natural language dialogue enables tutors to emphasize the co-construction of meaning -- seen in a small number of previous-generation systems such as AutoTutor and Watson Tutor \cite{graesser1999autotutor,afzal2019personality}. Conversations provide space for clarifying concepts, repairing misunderstandings, and drawing connections across ideas. This reflects long-standing evidence that rich discussions are central to learning, not only for delivering information but also for enabling students and tutors to check shared understanding and refine conceptual knowledge. Conversational tutoring thus opens the possibility of centering learning interactions on meaning rather than solely on procedures.

A second opportunity is to center student agency and micro-agency. Traditional ITS allowed limited forms of choice (e.g., problem selection), whereas conversational tutors can support more flexible and student-driven pathways. Students can initiate questions, explore alternative solution strategies, or redirect the focus of a session in ways that reflect their interests and goals. Even at the micro level, small conversational moves — such as deciding how to proceed in a problem or how to phrase an explanation — provide opportunities for agency. These forms of agency are associated with higher motivation and engagement, and conversational tutoring makes it feasible to embed them directly into instructional interactions.

Finally, conversational tutoring may allow for richer diagnosis of students' reasoning through seamless formative assessment. Large language models, when integrated with structured knowledge representations, may be able to identify not just whether an answer is correct but why a student responded in a particular way. For example, the details of a student's answer can indicate why their answer was wrong \cite{Scarlatos2025}, or tutorial dialogue can be used to follow-up an error with diagnosis questions \cite{graesser1999autotutor,afzal2019personality} showing promise for using dialogue to assess conceptual understanding, map skills, and detect misconceptions in real time. This capacity would move beyond traditional knowledge tracing by situating diagnosis within the unfolding conversation, allowing tutors to respond to the reasoning behind an answer rather than only to its outcome.

In sum, conversational tutoring makes it possible to center aspects of learning that earlier systems could only approximate. By prioritizing dialogue for meaning-making, supporting authentic student agency, and diagnosing reasoning through conversation, the field can expand what tutoring systems are designed to achieve.

\section{Study: Student Engagement and Learning Outcomes from Conversational AI Tutoring}
If conversational tutoring is to fulfill its potential, several areas of research must be advanced in parallel. At present, most prototypes rely on default large language model behaviors, which are often misaligned with evidence-based tutoring practices. This gap highlights the need for design research that integrates techniques such as prompt engineering, retrieval-augmented generation, and hybrid architectures to ensure that conversations reflect established pedagogical strategies rather than response strategies optimized for a ``helpful assistant.''
Equally pressing is the need for rigorous evaluations of efficacy. Many of the current claims about conversational tutors are based on pilot studies or anecdotal classroom reports. To establish confidence in these systems, the field will need large-scale randomized trials, controlled A/B tests of design features, and longitudinal studies that examine not only immediate learning gains but also persistence, transfer, and equity effects.

Research must also attend to the student experience of conversational tutoring. One open question is how students relate to AI tutors over time: whether they see them as helpful tools, responsive partners, or something in between. Studies suggest that relationship building is an essential component to tutoring \cite{Juel1996,KraftFalken2021}. Early studies on the relational aspects generative AI show that these models contain empathy capacity for interpersonal closeness that may outperform humans \cite{kleinert2026ai}, which may have concerned effects on human emotions and behavior \cite{ho2025potential}. However, it remains unclear whether students can or should form sustained relationships with AI tutors, and what the benefits or risks of such relationships might be. Investigating this dynamic is crucial, as tutoring is not only instructional but also relational.

Finally, conversational tutors must be understood as part of a larger human–AI ecosystem. Teachers, parents, and peers will continue to play central roles in learning, and the most effective systems will likely be those that keep humans informed and empowered. Research should explore how AI tutors can provide actionable insights to teachers, when a shift to human support or interaction is warranted, and how responsibility for instructional decisions can best be shared between human and AI partners.

In short, advancing conversational tutoring requires more than technical improvement. It calls for a coordinated research agenda that addresses design, evaluates efficacy, examines student experience, and integrates human collaboration. Only through such a program can we understand not just what conversational tutors can do, but how they can meaningfully support learning at scale.

\section{Example Implementation: Inner-Loop Response to Incorrect Answer}

In order to display how legacy systems and generative AI might interact to create a conversational AI tutor interaction, we present one (non-comprehensive) example in Figure \ref{fig:diag} to operationalize how a conversational AI tutor could respond to a student’s incorrect answer. Critically, it demonstrates the two-fold value of legacy ITS identifying students' cognitive and affective states for LLMs to respond.  In this design, legacy ITS components first process student responses and interaction traces to estimate learner-state parameters. A knowledge tracing model produces probabilities that the error reflects a slip versus non-mastery. When slip probability is high, this signal conditions the LLM to generate a light-touch response, such as encouragement plus a self-check prompt (e.g., ``You’re very close — what would you change?'').

If the mastery probability is low, the interaction may be routed toward probing the student further. The LLM is prompted to ask a short sequence of diagnostic questions to elicit the student’s reasoning, which informs misconception mapping (e.g., ``Can you walk through how you got that answer?''). Misconception estimates, combined with mastery probabilities, then determine the remediation strategy.

When multiple misconceptions are likely, and the likelihood of mastery is low, the LLM is prompted to deliver stronger remediation, such as a worked example followed by heavily scaffolded guidance through a parallel problem. If misconceptions are based in a need for more prerequisite knowledge, the system may even route a student to new content that address those needs before bringing a student back to the current problem  When mastery is moderate and likely misconceptions are fewer, the LLM is prompted to provide targeted hints and lighter scaffolding instead. In this way, structured learner-model estimates map directly to differentiated conversational paths, while the generative model is used to realize the selected pedagogical move in natural language.

Over time, these components are likely to become more tightly integrated and mutually informative. For example, misconception mapping may increasingly draw on evidence elicited through probing dialogue, rather than relying solely on performance traces as in earlier response-only approaches \cite{Liu_Patel_Koedinger_2016}. Recent work has already begun to use LLMs to help construct and refine knowledge spaces \cite{LearningCommons2024} and to support dialogue-informed knowledge tracing models \cite{Scarlatos2025}. However, these hybrid methods remain early-stage and require further validation to ensure that generative inferences about learner knowledge are reliable, stable, and instructionally actionable. This trajectory underscores a central theme of this paper: the best potential comes not from replacing established ITS methods with conversational AI, but from systematically integrating them.

\section{Conclusion}
It is nearly inevitable that the next generation of ITS will look vastly different from the previous generation. For decades, ITS research has demonstrated that technology can approximate some of the benefits of human tutoring. Generative AI now introduces the possibility of tutoring that unfolds in natural language dialogue, more closely resembling the conversations that make human tutoring so powerful. Yet there is little evidence that generative AI alone can match the multifaceted nature of high-quality tutoring, which requires simultaneously predicting, understanding, and responding to a multiplicity of students' emotional and cognitive needs. Thus, it is important that we keep proven methods that make sense of students' work so that AI tutors can adapt responsively. However, we must lean into AI's ability to generate responses and content based on student needs. This will allow the next generation of systems to center dialogue as a space for meaning-making, elevating dimensions that earlier systems could not support. All of this will require extensive research into design alignment, and efficacy will be essential for building reliably effective and safe systems. By integrating prior research with the unique affordances of generative AI, the field can move closer to the goal of universal access to high-quality tutoring — tutoring that is not only scalable, but also conversational, responsive, and student-centered in design.

\bibliographystyle{ACM-Reference-Format}
\bibliography{biblography}

\end{document}